\documentstyle[prl,epsfig,aps]{revtex}
\begin{document}
\preprint{LPTENS-97/18}
\twocolumn[\hsize\textwidth\columnwidth\hsize\csname@twocolumnfalse\endcsname

\title{Hole-burning experiments within solvable glassy models}

\author{Leticia F. Cugliandolo$^*$ and 
Jos\'e Luis Iguain$^{**}$ }
\address{ $^*$
\it Laboratoire de Physique Th\'eorique de l'\'Ecole Normale 
Sup\'erieure,
\cite{add3} 
24 rue Lhomond, 75231 Paris Cedex 05, France and \\
 Laboratoire de Physique Th\'eorique  et Hautes Energies, Jussieu, 
5\`eme \'etage,  Tour 24, 4 Place Jussieu, 75005 Paris France
\\
$^{**}$
\it Departamento de F\'{\i}sica, Universidad Nacional de Mar del Plata, 
De\'an Funes 3350, 
7600, Mar del Plata, Argentina}

\date\today
\maketitle
\begin{abstract}
We reproduce the results of non-resonant spectral hole-burning experiments  
with fully-connected (equivalently infinite-dimensional) glassy models 
that are generalizations of the mode-coupling approach 
to nonequilibrium situations. 
We show that an ac-field modifies the integrated linear response and the 
correlation function in a way that depends on the amplitude and frequency of 
the pumping field. We study the effect of the waiting and 
recovery-times and the number of oscillations applied. This calculation 
will help descriminating which results can and which cannot be attributed to 
dynamic heterogeneities in real systems.
\end{abstract}
\vspace{.2cm}
\hspace{1.8cm} PACS Numbers: 64.70.Pf, 75.10Nr
\twocolumn 
\vskip.5pc]
\narrowtext

One of the most interesting questions in glassy physics is whether
{\it localized spatial heterogenities} are generated 
in supercooled liquids and glasses. \cite{Sillescu}

In most supercooled liquids, the linear response to small external 
perturbations is nonexponential in the time-difference $\tau$. 
Within the ``heteregeneous scenario'',
the stretching is due to the existence of dynamically
distinguishable entities in the sample,
each of them relaxing exponentially with its
own characteristic time. 
A different interpretation is that the macroscopic response is
intrinsically nonexponential. 
In the glass phase, the relaxation is nonstationary and the dependence
in $\tau$ is also much slower than exponential. 

The heterogeneous regions, if they exist, 
are expected to be nanoscopic.
The development of experimental techniques capable of giving evidence  
for the existence of such distinguishable spatial regions 
has been a challenge for experimentalists. 

\begin{figure}
\centerline{\hbox{
   \epsfig{figure=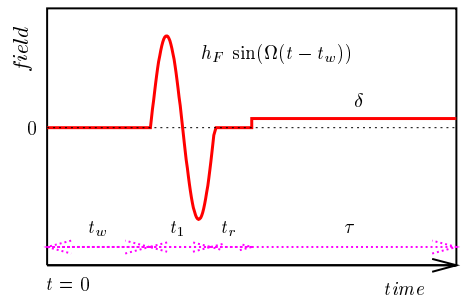,width=7cm}}
 }
\caption{Wait, pump, recovery and probe scheme. 
}
\label{schema}
\end{figure}

With  non-resonant spectral hole-burning (NSHB) 
techniques one expects to  probe,
selectively, the microscopic responses.\cite{Bohmer1} 
The method is based on a wait, pump, recovery and 
probe scheme depicted in Fig.~\ref{schema}.
The amplitude of the ac perturbation is  
sufficiently large to pump energy in the sample,  
modifying the response as a 
linear function of the absorbed energy.
The step-like perturbation $\delta$ is very weak and serves 
as a probe to measure the integrated linear response of the full system.
The large ac and small dc fields can be 
magnetic, electric, or any 
other perturbation relevant for the sample studied. 
The idea behind the method is that the comparison of the 
modified (perturbed by the oscillation) and unmodified
(unperturbed) integrated responses yield information 
about the microscopic structure of the sample.
On the one hand, a spatially homogeneous sample will  absorb energy uniformly 
and its modified integrated response
is expected to be a simple translation towards shorter time-differences $\tau$ 
of the unmodified one. On the other hand, in a heterogeneous sample,
the degrees of freedom that respond near the pump frequency
$\Omega$ are expected to absorb an important amount of energy and a maximum 
difference in the relaxation (equivalently, a spectral
hole) is expected to generate around $t\sim 1/\Omega$.    

The NSHB technique has been first applied 
to the study of supercooled liquids.
The polarization response of dielectric samples, glycerol and 
propylene carbonate,  
was measured after being  modified by an ac electric field.\cite{Bohmer1}
More recently, ion-conducting glasses like CKN \cite{Bohmer2}, 
relaxor ferroelectrics (90PMN-10PT ceramics) \cite{Bohmer3}
and spin-glasses (5\% Au:Fe) \cite{Chamberlin}
were  studied with similar methods.    
The results have been interpreted 
as evidence for the existence of 
spatial heterogeneities.
We show here that their main features
can be reproduced by a system with {\it no spatial structure}. 
We use one model, out of  a family, that captures
many of the experimentally observed features of super-cooled 
liquids and glasses as, for instance,  
a two-step equilibrium relaxation close and above
$T_c$ \cite{Gotze}, aging effects below $T_c$ \cite{Cuku}, etc. 
The model is the $p$  spherical spin-glass \cite{Crso}, that is intimately related to the 
$F_{p-1}$ mode-coupling model \cite{Kith}. It can be interpreted as 
a system of $N$ fully-connected continuous spins
or as a model of a particles in an infinite
dimensional random environment. \cite{review}  
In both cases, no reference to a geometry in real space nor 
any identification 
of spatially distinguishable regions can be made. 

In the presence of a uniform field, the model  
is
\begin{equation}
H_J[{\bf s}] = \sum_{i_1 \leq \dots \leq i_p} J_{i_1\dots i_p} \; s_{i_1} \dots s_{i_p}
+
h \; \sum_{i=1}^N s_i 
\; .
\label{ham_pspin}
\end{equation}
The interactions 
$J_{i_1\dots i_p}$ are quenched independent random  variables taken from
a Gaussian distribution with zero mean and 
variance $[J^2_{i_1\dots i_p}]_J={\tilde J}^2 p!/(2 N^{p-1})$. 
$p$ is a parameter and we take $p=3$.
Hereafter $[\;\;]_J$ represents an average over $P[J]$ and $\tilde J=1$.
The continuous variables $s_i$ are constrained  spherically $\sum_{i=1}^N s_i^2=N$.
A stochastic evolution is given to ${\bbox s}$, 
$
\dot s_i(t) = -\delta_{s_i(t)} H_J[{\bbox s}] + \xi_i(t) 
$
with $\xi_i$ a white noise with $\langle \xi_i\rangle =0$ and  
$
\langle \xi_i(t)  \xi_i(t') \rangle = 2 T \delta(t-t') 
$. When $N\to \infty$, standard techniques
lead to a set of coupled integro-differential equations 
for the autocorrelation $
NC(t,t') \equiv \sum_{i=1}^N [\langle s_i(t) s_j(t') \rangle]_J
$ and the linear response 
$
R(t,t') \equiv \left. \sum_{i=1}^N 
\delta [\langle s_i(t) \rangle]_J /\delta
\delta_i(t')  \right|_{\delta=0}
$, with
$\delta_i(t')$ an infinitesimal
perturbation modifying the energy at time 
$t'$ according to $H \to H - \sum_i \delta_i s_i$. 
The dynamic equations read \cite{Cuku3}
\begin{eqnarray} 
& & \partial_t C(t,t')
=
 -z(t) \, C(t,t') +
\frac{p}{2} 
\int_0^{t'} dt''  C^{p-1}(t,t'')  R(t',t'') 
\nonumber\\
& & 
\;\;\;\;\;\;\;
+
\frac{p (p-1)}{2}\int_0^t dt'' C^{p-2}(t,t'') R(t,t'') C(t'',t')
\nonumber\\
& &
\;\;\;\;\;\;\;
+ 2T R(t',t) + h(t) \int_0^{t'} dt'' h(t'') R(t',t'')
\; ,
\label{eqC}
\\
& & 
\partial_t R(t,t') 
=
 -z(t) \, R(t,t')
\nonumber \\
& &   
\;\;\;\;\;\;\;+  \frac{p (p-1)}{2}
 \int_{t'}^t dt''  C^{p-2}(t,t'') R(t,t'')  R(t'',t')
\; ,
\label{eqR}
\end{eqnarray}
The Lagrange  multiplier  $z(t)$ enforces the spherical constraint
and an integral equation for it follows from Eq.~(\ref{eqC}) and 
the condition $C(t,t)=1$.
In deriving these equations, a random initial condition 
at $t_0=0$ has been used.
It corresponds to an  infinitely fast quench from equilibrium 
at $T=\infty$
to the working temperature $T$. The evolution continues in 
isothermal conditions. 

In the absence of energy pumping, these models have 
a dynamic phase transition at a ($p$-dependent) 
critical temperature $T_c$, $T_c \sim 0.61$ for $p=3$. 
 When an external ac-field is applied, it  drives 
the system out-of-equilibrium and stationarity and 
FDT do not necessarily  hold at {\it any} temperature. 
The question as to whether the clearcut dynamic transition survives 
under an oscillatory field is open and we do not address it here. 
We simply
study the dynamics close to the critical temperature in 
the absence of the field by
constructing a numerical solution to Eqs.~(\ref{eqC}) and (\ref{eqR}) 
with a constant grid algorithm of spacing $\epsilon$. 
We present data for small spacings, typically $\epsilon=0.02$, 
to minimize the numerical errors.
Due to the fact that Eqs.~(\ref{eqC}) and (\ref{eqR}) include integrals
ranging  from $t_0=0$ to present time $t$,
the algorithm is limited to a maximum number of iterations 
of the order of $8000$ that imposes a lower limit 
$\Omega \sim 2\pi/(8000 \epsilon) \sim 0.1$
to the frequencies we use.

A word of caution concerning the scheme in 
Fig.~\ref{schema} and the times involved 
is in order. For the purpose of collecting the data for each reference  
unmodified integrated response, 
the sample is prepared at the working temperature $T$ at $t_0=0$ and let freely evolve during a total waiting time $t_w+t_1+t_r$. Depending on $T$, 
this interval may or may not be enough to equilibrate the sample.
($t_1$ is chosen as $t_1=2\pi  n_c/\Omega$ with 
$\Omega$ the angular velocity of 
the field that will be used to record the modified curve.) 
A constant infinitesimal probe $\delta$ is applied after $t_w+t_1+t_r$
to measure
\begin{eqnarray*}
\Phi(\tau) \equiv \int_0^\tau d\tau' \, R(t_w+t_1+t_r+\tau, t_w+t_1+t_r+\tau')
\; .
\end{eqnarray*}
As an abuse of notation we explicitate only the $\tau$ dependence
and eliminate the possible $t_w+t_1+t_r$ dependence.
The modified integrated response $\Phi^*$ is measured after
waiting $t_w$, applying $n_c$ oscillations of duration 
$t_1=2\pi n_c/\Omega$, further waiting $t_r$, and only then 
applying the probe $\delta$. 
The effect of the ac perturbation is then quantified by 
studying the difference:
\begin{equation}
\Delta \Phi \equiv \Phi^* - \Phi
\; .
\end{equation}  

\begin{figure}
\centerline{\hbox{
   \epsfig{figure=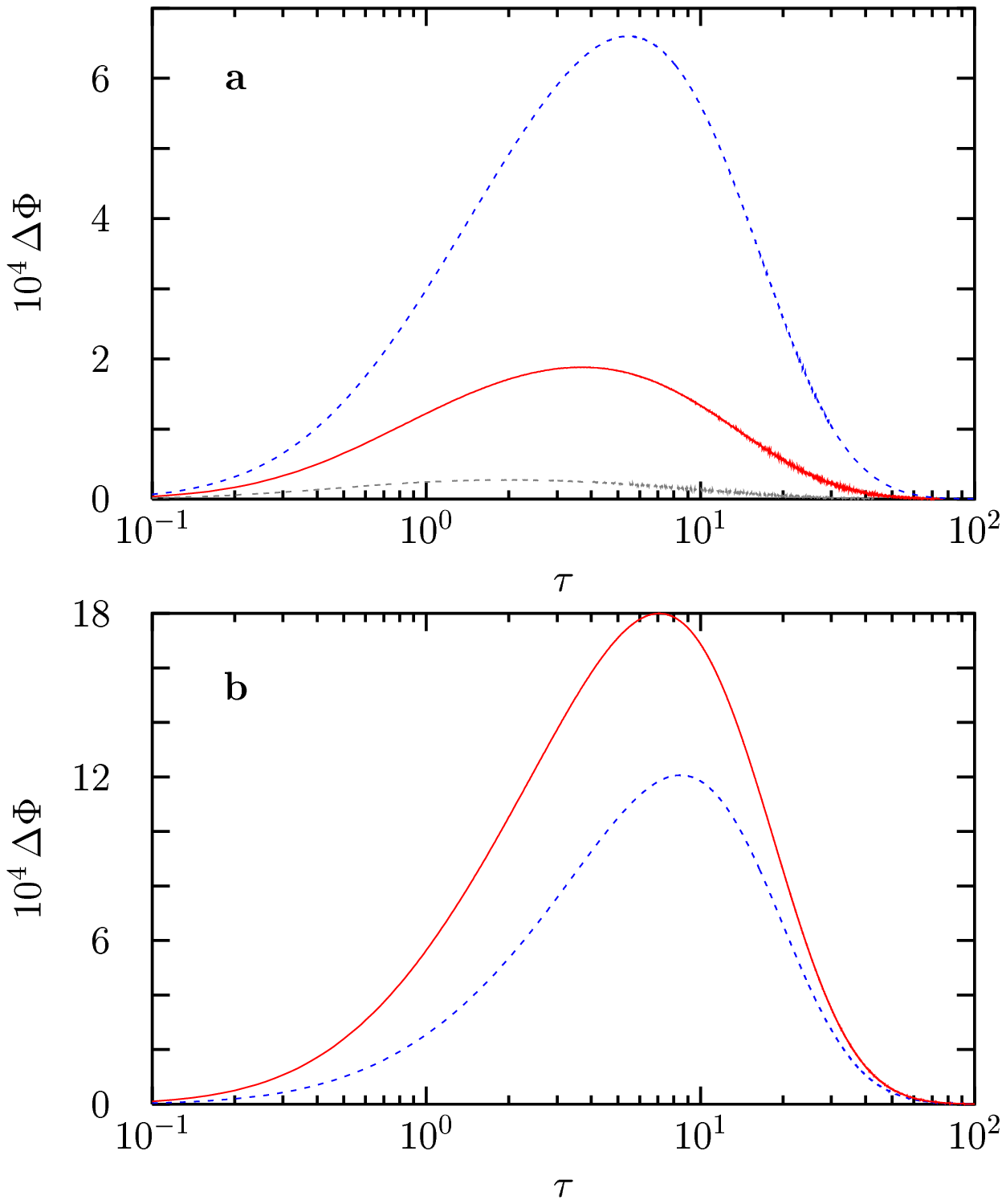,width=8cm}}
 }
\caption{
Time-difference dependence of the distortion 
$\Delta\Phi$ due to a single oscillation in a 
log-linear scale. $T=0.8> T_c$, $h_F=0.1$, $t_w=t_r=0$. 
At high pumping frequencies $\Omega > \Omega_c\sim 1$, shown in panel a, 
both the height of the peak $\Delta\Phi_m$ and its position $t_m$
decrease with increasing frequency. 
The dotted (blue), solid (red) and dashed (black) 
curves correspond to $\Omega=1,2,5$ respectively. 
In panel b, $\Omega < \Omega_c\sim 1$ and $\Delta\Phi_m$
increases with increasing $\Omega$ while $t_m$ is almost unchanged. 
The dotted (blue) and solid (red) curves correspond to $\Omega=0.1$ and $0.2$, 
respectively. }
\label{Delta_Phi_1}
\centerline{\hbox{
   \epsfig{figure=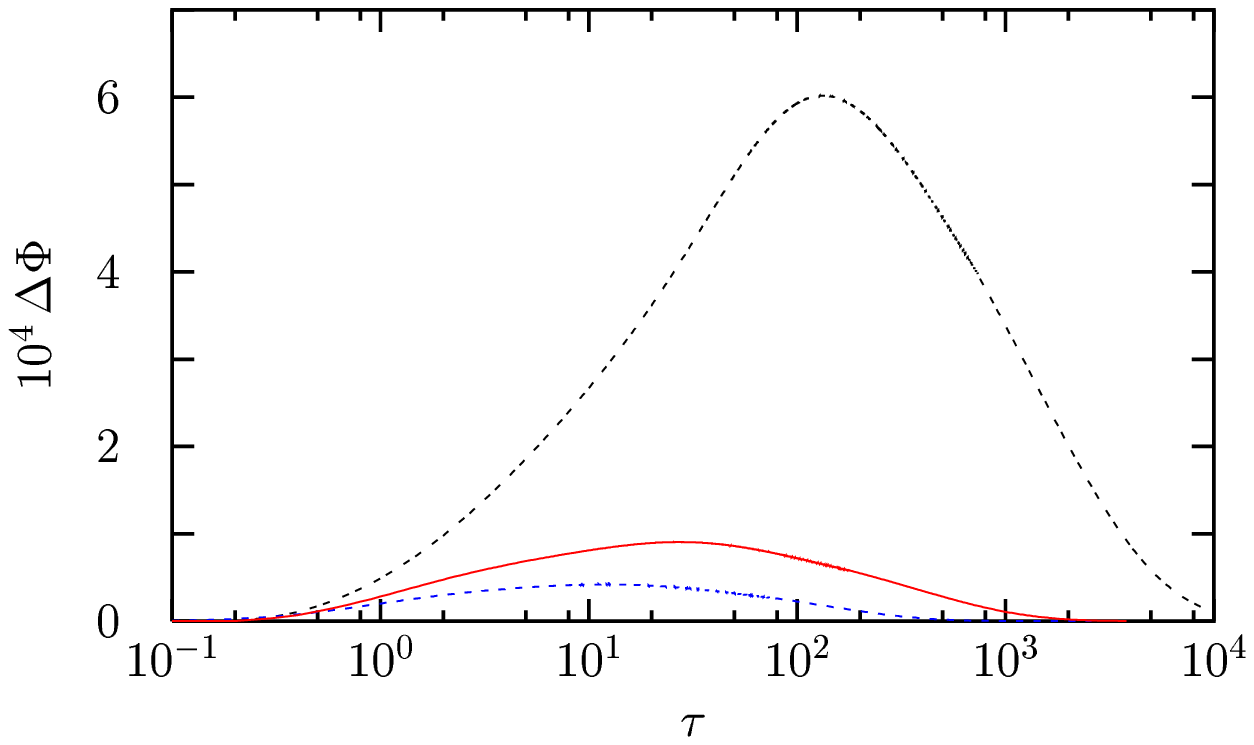,width=8cm}}
 }
\caption{
The same plot as in Fig.~\ref{Delta_Phi_1} for  $T=0.59 < T_c$.
For all pumping frequencies explored, the peak moves towars shorter 
times and its height decreases for increasing frequencies. The dashed (black), 
solid (red) and dotted (blue) curves correspond to $\Omega=0.1,0.5$ and $1$,
respectively.}
\label{Delta_Phi_2}
\end{figure}

We have examined $\Delta \Phi$ at $T=0.8 > T_c$ and $T=0.59< T_c$.
We pump one oscillation with $h_F=0.1$
and later check that this field is small enough to provoke a 
spectral modification that is linear in the absorbed energy (see
Fig.~\ref{checklinearity} below).  
For simplicity, we start by choosing $t_w=t_r=0$. 
In Fig.~\ref{Delta_Phi_1} we show  $\Delta \Phi$ against $\log \tau$ 
for different  $\Omega$ at $T=0.8$.
All the curves are bell-shaped and vanish both at short and long times.  
In panel a, the $\Omega$s are larger than a threshold value
$\Omega_c \sim 1$. The height of the
peak $\Delta\Phi_m \equiv \max ( \Delta\Phi)$ 
decreases with increasing frequency reaching the limit $\Delta\Phi_m= 0$
for $\Omega\to\infty$. In addition, the location of the peak 
$t_m$ moves towards longer times when $\Omega$ decreases. 
In panel b,  $\Omega< \Omega_c$ and the behaviour of the 
height of the peak is the opposite, it decreases when $\Omega$ decreases and, 
within numerically errors, 
its position is either independent of $\Omega$ or it very 
smoothly moves towards shorter times for increasing $\Omega$. 
The nonmonotonic behaviour of $\Phi_m$ with $\Omega$
is a consequence of the interplay between $t_\alpha$,
the $\alpha$ relaxation time, and $2\pi/\Omega$ the period of 
the oscillation. The term $\int_0^{\min(t',t_1)} dt'' h(t'') R(t',t'')$
in Eq.~(\ref{eqC}) controls the effect of the field and, clearly, 
vanishes in the limits $\Omega\to\infty$ and $\Omega\to 0$. 
The inversion then occurs
at a frequency $\Omega_c$  that is of the order of $2\pi/t_\alpha$.
These results qualitatively coincide with the measurements 
of the electric relaxation
in CKN at $T<T_g$ in Fig~1 a and b of Ref.~\cite{Bohmer2}. 
In Fig.~\ref{Delta_Phi_2} we show $\Delta \Phi$ against 
$\log \tau$ for different $\Omega$ 
at $T=0.59$. For all $\Omega$ we reproduced the 
situation of panel a in Fig.~\ref{Delta_Phi_1}, as if $\Omega > \Omega_c$. 
We have not found a threshold $\Omega_c$, that has gone 
below the minimum $\Omega$ reachable with the algorithm.

\vspace{-.25cm}

The maximum modification of the relaxation $\Delta\Phi_m$ 
increases quadratically with the square of the amplitude of the pumping field 
$h_F$, and hence linearly in the absorbed energy,  
as long as $h_F \leq 1$. In Fig.~\ref{checklinearity} we display the 
relation $\Delta\Phi_m\propto h_F^2$ in a log-log scale for the two 
temperatures explored. The amplitude $h_F=0.1$ used in Figs.~\ref{Delta_Phi_1} 
and \ref{Delta_Phi_2}  is in the linear regime.   

\begin{figure}
\centerline{\hbox{
   \epsfig{figure=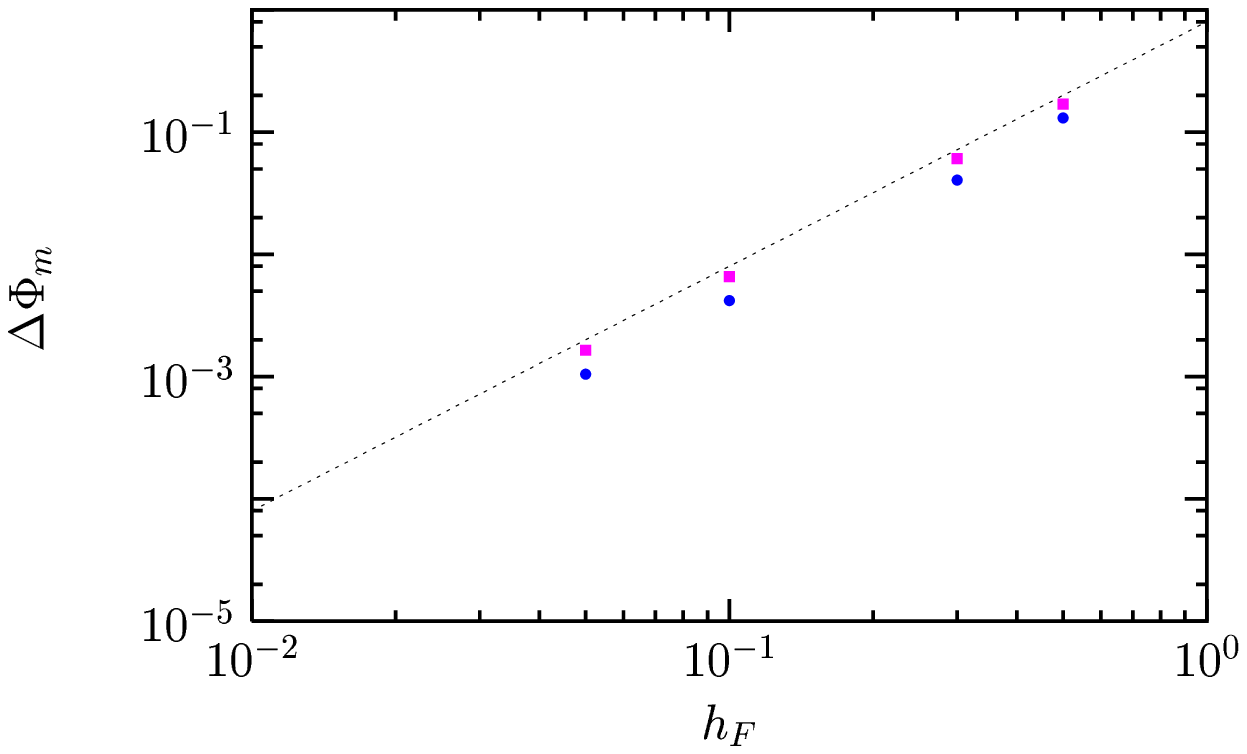,width=8cm}}
 }
\caption{
Check of $\Delta\Phi_m\propto h_F^2$ in a log-log scale.
Squares (pink) correspond to $T=0.8$ and circles (blue) 
to $T=0.59$. In both cases
one cycle of an ac-field with $\Omega=1$ was
applied. The line has a slope equal to $2$ and is a guide-to-the-eye. 
The linear relation breaks down beyond $h_F \sim 1$.
}
\label{checklinearity}
\centerline{\hbox{
   \epsfig{figure=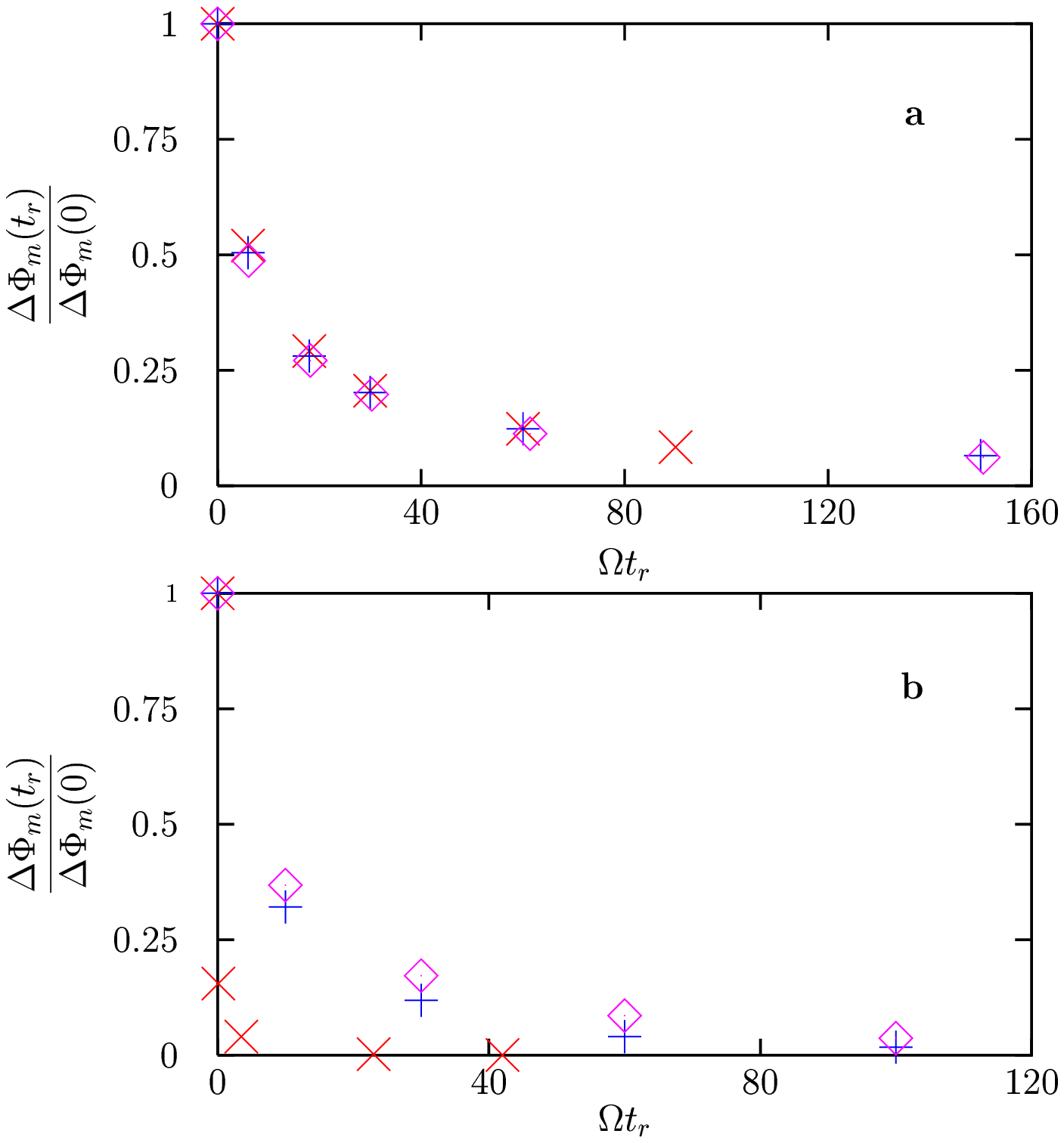,width=8cm}}
 }
\caption{
 The normalized maximum distortion for several 
recovery times $t_r$. In panel a $T=0.59$ and 
crosses, diamonds and pluses correspond
to $\Omega=1,2$ and $3$, respectively.
In panel b $T=0.8$ and
crosses, pluses and diamonds correspond
to $\Omega=1,5,10$, respectively.}
\label{recovery}
\end{figure}

The effect of the pump diminishes with increasing recovery time $t_r$.
A convenient way of displaying this result is to plot the normalized
maximum deviation $\Delta\Phi_m(t_r)/\Delta\Phi_m(0)$ {\it vs} $\Omega t_r$.
Using several frequencies and recovery times, 
we verified that this scaling holds for $T=0.59$ but does not hold 
for $T=0.8$, as shown in Fig.~\ref{recovery}. This simple saling 
holds very nicely in the relaxor ferroelectric \cite{Bohmer3}
and in the spin-glass \cite{Chamberlin} but it is very different 
from the $\Omega$-independence of the propylene carbonate \cite{Bohmer1}. 

\begin{figure}
\centerline{\hbox{
   \epsfig{figure=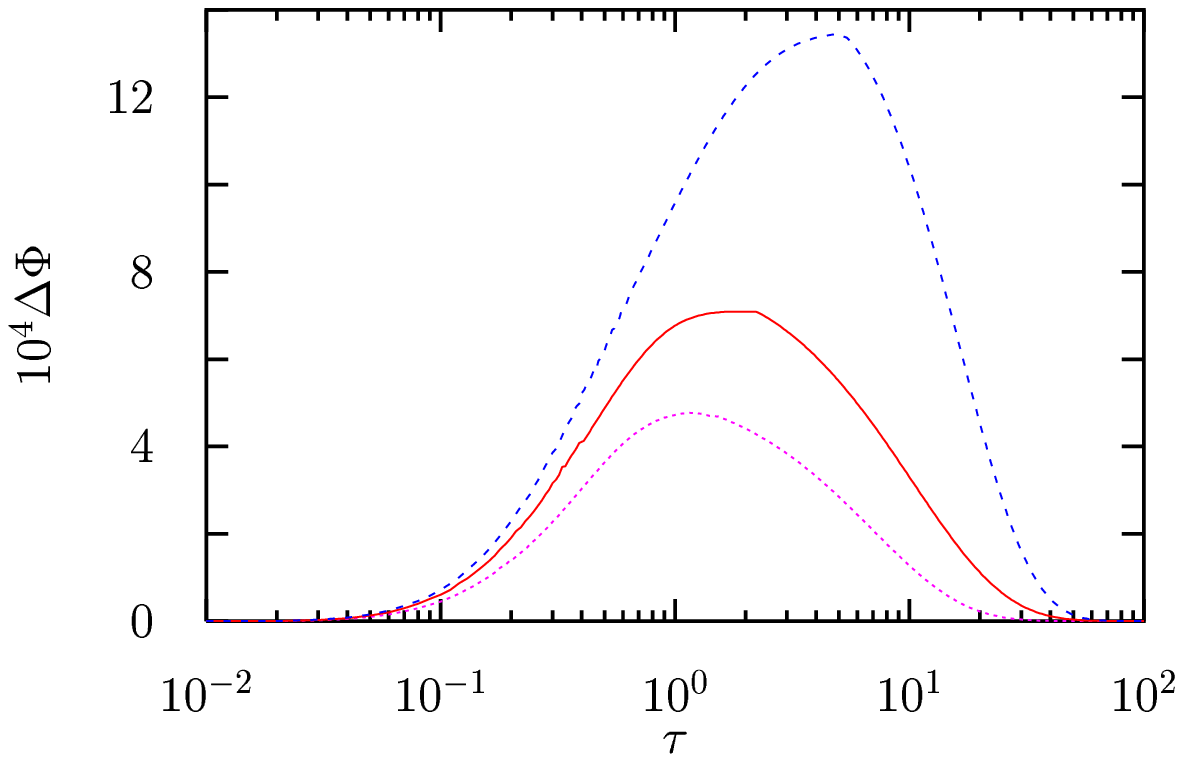,width=8cm}}
 }
\caption{
Effect of several cycles  at 
$T=0.8$ for $\Omega=10$. The dashed (blue), solid (red) and 
dotted (pink) curves correspond to 
$n_c=10,2,1$ respectively. 
}
\label{ncyc}
\centerline{\hbox{
   \epsfig{figure=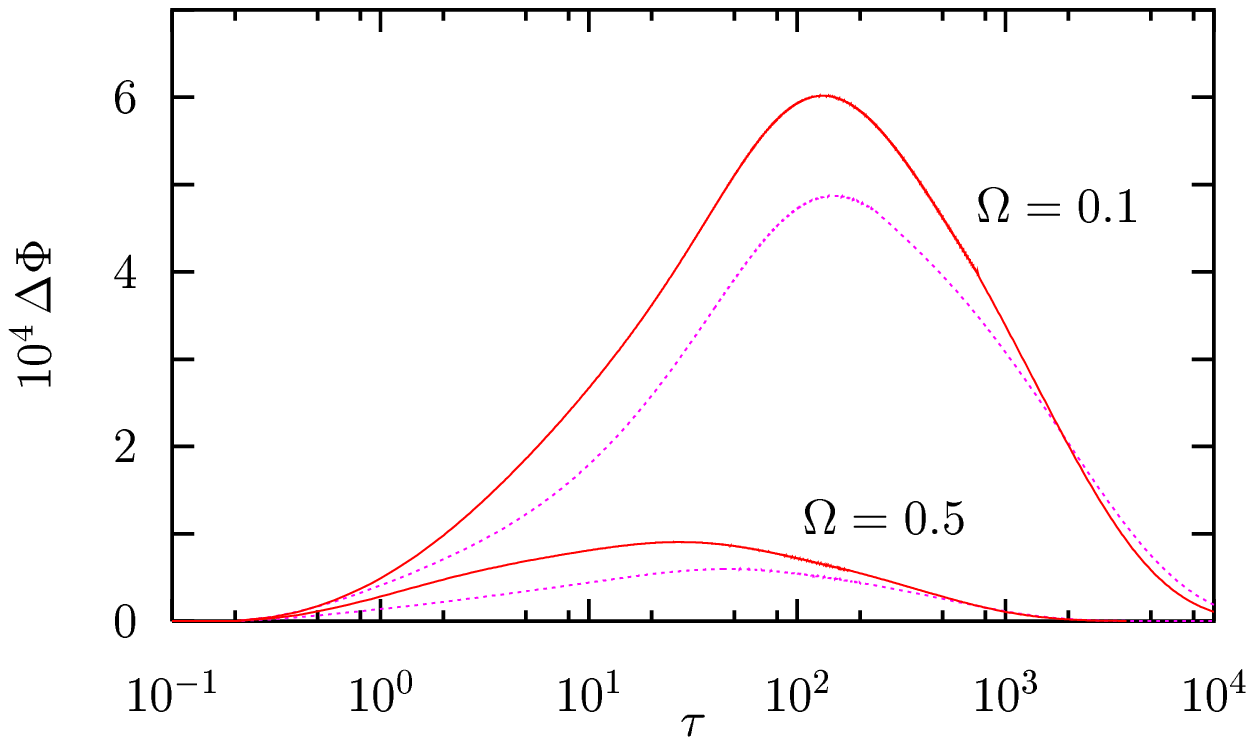,width=8cm}}
 }
\caption{
Distortion for $t_w=8$ with 
dots (pink) compared with the one for $t_w=0$ with full line (red)
at $T=0.59$.
}
\label{tw-dep}
\end{figure}

Up to now, the effect of a single cycle of different frequencies 
has been studied. Another procedure can be envisaged.
Since $t_1=2\pi n_c/\Omega$, we can change $t_1$ 
by applying different numbers of cycles $n_c$ while keeping $\Omega$ fixed. 
In Fig.~\ref{ncyc} we show the distortion due to 
$n_c=10,2,1$ cycles with $\Omega=10$ at $T=0.8$.
The qualitative dependence on $n_c$ is indeed the same as 
the dependence on $1/\Omega$:
the peaks are displaced towards longer times with increasing $n_c$
(longer $t_1$).
This behaviour is similar to the results
obtained for propylene carbonate in Fig.~11 of 
Ref.~\cite{Bohmer1}b. though we do not reach the expected saturation within 
our accessible time window.

Below $T_c$ the nonperturbed model never equilibrates and the 
relaxation depends on $t_w$. Indeed, $t_\alpha$ is an approximately 
linear function of $t_w$\cite{Cuku,review} and the distortion 
might depend on $t_w$. We compare $\Delta\Phi$ vs $\log\tau$ 
for two $t_w$'s in Fig.~\ref{tw-dep}. 
 
\begin{figure}
\centerline{\hbox{
   \epsfig{figure=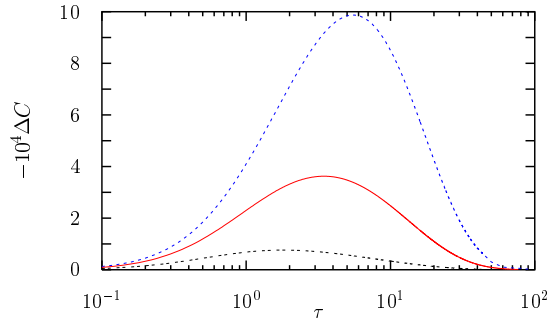,width=8cm}}
 }
\caption{
Change in the autocorrelation. $T=0.8$ and $\Omega=1$ (dashed blue), 
$\Omega=2$ (solid red) and $\Omega=5$ (dotted black).
}
\label{corr}
\end{figure}

Finally, we checked that the effect of one 
or many pump oscillations 
on the difference $\Delta C 
\equiv C^*(t_w+t_1+t_r+\tau,t_w+t_1+t_r)-C(t_w+t_1+t_r+\tau,t_w+t_1+t_r)$ 
is very similar to the one observed in $\Delta\Phi$. 
This observation is interesting since it is easier to compute numerically
correlations than responses. 
Figure~\ref{corr} shows the modification observed at 
$T=0.8$ and $\Omega > \Omega_c$ (to be compared to Fig.~\ref{Delta_Phi_1}).

We conclude by stressing that we
do not claim that spatial  heterogeneities do not exist 
in real glassy systems. We just wish to stress that the 
ambiguities in the interpretation of
experimental results have to be eliminated in order to have 
unequivocal evidence for them. The detailed comparison of the 
experimental measurements to the behaviour of glassy models 
{\it with and without space} will certainly help us refine the experimental 
techniques. Numerical simulations can play an important role in this 
respect. 

LFC and JLI thank the Dept. of Phys. (UNMDP) and  LPTHE (Jussieu) 
for hospitality, and ECOS-Sud, CONICET and UNMDP for financial support. 
We thank R. B\"ohmer, H. Cummins, G. Diezemann, M. Ediger, 
J. Kurchan and G. Mc Kenna for very useful discussions and 
T. Grigera, N. Israeloff and E. Vidal-Russel for introducing 
us to the hole-burning experiments.

\end{document}